\documentclass[11pt]{article}
\usepackage[a4paper, total={6in, 9.5in}]{geometry}
\usepackage[T1]{fontenc}
\usepackage[utf8]{inputenc}
\usepackage{enumitem}
\usepackage{amsmath,amssymb}
\usepackage{amsthm} 
\usepackage{graphicx}
\usepackage[toc,page]{appendix}
\usepackage{float}
\usepackage{hyperref}
\usepackage{xcolor}
\hypersetup{
    colorlinks,
    linkcolor={red!50!black},
    citecolor={blue!50!black},
    urlcolor={blue!80!black}
}

\setlist[itemize]{label=$\cdot$}
\usepackage[english]{babel}
\usepackage{biblatex}
\usepackage{indentfirst}
\usepackage[english]{babel}
\usepackage{csquotes}
\usepackage{color, colortbl}
\definecolor{Gray}{gray}{0.9}
\definecolor{LightCyan}{rgb}{0.88,1,1}
\usepackage[first=0,last=9]{lcg}

\newcolumntype{b}{>{\columncolor{LightCyan}}c}
\newcolumntype{d}{>{\columncolor{Apricot}}c}
 
\newtheorem{theorem}{Theorem}[section]
\newtheorem{corollary}{Corollary}[theorem]
\newtheorem{lemma}[theorem]{Lemma}

\usepackage{algorithm}
\usepackage{algorithmic}

\def\b{\ensuremath\beta}

\def\D{\ensuremath\Delta}
\def\k{\ensuremath\kappa}
\def\g{\ensuremath\gamma}

\def\p{\ensuremath\phi}

\def\eps{\ensuremath\epsilon}

\def\s{\ensuremath\sigma}

\newcount\Comments  

\newcommand{\kibitz}[2]{\ifnum\Comments=1{\color{#1}{#2}}\fi}

\usepackage{sgame, tikz} 
\usetikzlibrary{trees, calc} 
\usetikzlibrary{decorations.pathreplacing}

\newcommand{\floor}[1]{\left\lfloor #1 \right\rfloor }

\newcommand{\even}[1]{\textrm{even}(#1)}
\newcommand{\odd}[1]{\textrm{odd}(#1)}

\usepackage{booktabs} 
    
\makeatletter
\def\and{%
  \end{tabular}%
  \hskip 0.4em \@plus.17fil\relax
  \begin{tabular}[t]{c}}
\makeatother

\author{Daniel J. Moroz\thanks{Email: \texttt{dmoroz@g.harvard.edu}}\\Harvard University\\
\and Daniel J. Aronoff\thanks{Email: \texttt{\{daronoff,narula\}@mit.edu}}\\MIT Economics
\and Neha Narula\footnotemark[2]\\MIT Media Lab
\and David C. Parkes\thanks{Email: \texttt{parkes@eecs.harvard.edu}}\\Harvard University}

\date{February 2020}

\begin{document}

\title{\textbf{Double-Spend Counterattacks: Threat of Retaliation in Proof-of-Work Systems}}

\maketitle

\begin{abstract}
Proof-of-Work mining is intended to provide blockchains with robustness against double-spend attacks. 
However, an economic analysis that follows from Budish (2018), 
which considers free entry conditions together with the ability to rent sufficient hashrate to conduct an attack,
suggests that the resulting block rewards can make an attack  cheap.
We formalize a defense to double-spend attacks.
We show that when the victim can counterattack in the same way as the attacker, this leads to a variation on the classic game-theoretic \textit{War of Attrition} model.  The threat of this kind of counterattack induces a subgame perfect equilibrium in which no attack occurs in the first place. 
\end{abstract}

\section{Introduction}

Bitcoin is designed to solve the challenge of reaching agreement (consensus) over an ordered list of transactions (a blockchain) in a permissionless, peer-to-peer electronic cash system~\cite{Nakamoto}. 
It does so by requiring participants to demonstrate computational activity, Proof-of-Work (PoW), and to build on the valid blockchain with the most work (heaviest-chain rule).\footnote{This is sometimes referred to as the longest chain rule, but it is in fact the chain with the most Proof-of-Work as measured by the software, hence heaviest-chain rule, and not the chain with the most blocks.}
In Bitcoin and other PoW-based cryptocurrencies, miners perform expensive computational work committing to a particular transaction history. 
Miners are incentivized to do so because they receive rewards that are only valid if the chain on which they mine is the chain accepted by other participants.

In the course of a \textit{double-spend attack}, an attacker rewrites a portion of the blockchain transaction history, spending the same token in two different ways. 
In this work we focus on the subset of double-spend attacks caused by obtaining a majority of the hashpower, known as \emph{51\% attacks}.\footnote{We do not consider zero-confirmation attacks or attacks due to software version incompatibility.} 
To conduct a double-spend attack, an attacker $A$ would create a blockchain transaction giving tokens to a victim, for example a cryptocurrency exchange. After an escrow waiting period, $A$ would exchange the tokens for a good, for example by selling the tokens for USD on the exchange. Meanwhile, $A$ would privately work to create an alternate chain history (a \textit{chain reorganization}, or \textit{reorg}) in which the original transaction is replaced by a self-payment, rendering the original transaction invalid (to do this, $A$ would need to complete more PoW on the alternate chain than the network had completed on the original chain). Upon receiving the USD, $A$ would reveal this new chain to the network, and all other participants faithfully following the heaviest-chain rule would switch to this rewritten history of transactions.

In fact, an economic analysis provided by Budish (2018)~\cite{Budish} suggests that, given a liquid market for hashrate, the cost of acquiring a majority of mining power on a chain for long enough to execute a double-spend attack would be small or negligible (see Section \ref{Budish-Section} for a modified version of this argument) because the attacker recoups the cost of attack in block rewards. The implication is that, with sufficient liquidity, many transactions completed on current PoW systems would be insecure against double-spend attacks. 
Moreover, markets for hashrate do exist. Nicehash is a marketplace that connects sellers and buyers of hashrate for different algorithms, and the website Crypto51 shows to what extent various chains are vulnerable to 51\% attacks through NiceHash~\cite{NiceHash, crypto51}.

\paragraph{Our Contributions.} We investigate the question of double-spend attacks through theoretical modeling and analysis, while also seeking support from empirical evidence. 

As our primary contribution, we provide a formal model of a {\em Retaliation game}, extending Budish's model~\cite{Budish} with a novel action (the victim's counterattack). We show in this model, where both the attacker and defender can procure hashrate, that double-spending is not profitable. 
We consider a variation on the widely studied {\em War of Attrition} model from game theory, and show a subgame perfect equilibrium in which no double-spend attack occurs even if a single attack is cheap relative to the contested transaction. Our result is achieved under mild assumptions, namely that the net cost of each successive attack increases over time, and the victim, for example an exchange, pays a reputation cost for being double-spent. 

Second, we modify Budish's model to show that, without considering counterattacks, double-spend attacks may be either free, cheap or impossible, depending on the market availability of hashrate and the price impact of an attack.\footnote{Auer~\cite{Auer} shows a similar result, independently and contemporaneously, though without connecting directly to the Budish model.}

Third, we compile known data on  double-spend attacks to show that attacks have occurred on low-PoW chains such as Bitcoin Gold but not high-PoW chains such as Bitcoin. As of February 2020, we also identify preliminary evidence of retaliations to double-spend attacks on Bitcoin Gold.

\medskip
\medskip

The rest of this paper is organized as follows: In Section \ref{Budish-Section} we introduce a modified version of the economic model introduced in Budish, adopting a different duration of attack, which changes the rental cost of attack and the block rewards that accrue from an attack. In Section \ref{Our-Model}, we introduce the Retaliation game, which considers the possibility of retaliation by the victim, and develop a subgame perfect equilibrium  analysis of no-attack. In Section \ref{Empirical-Investigations}, we present empirical results that show the existence of double-spend attacks on low-PoW chains but not high-PoW chains. In Section \ref{Discussion} we provide a discussion of the empirical results we see and their connection with our models. In Section~\ref{Conclusion} we conclude, and discuss directions for future work. 

\subsection{Related Work}

\subsubsection{Game-Theoretic Analyses of PoW Systems}

There are several studies analyzing game-theoretic models of PoW mining and double-spending specifically, with early work assuming that the attacker would never achieve a majority of the hashrate. 
In the original Bitcoin paper, Nakamoto (2008) \cite{Nakamoto} shows that the probability that a minority attacker would be able to double-spend the Bitcoin network decreases as the number of blocks found since a transaction increases. 
Rosenfeld (2012) \cite{Rosenfeld-DS} expands on these calculations, producing tables of the probability of successful attack and largest safe transaction size (in BTC), both as function of a potential attacker's hashrate ($<50\%$) and the number of blocks already mined on top of the transaction in question.

As we elaborate on in Section~\ref{Budish-Section}, our work is related to Budish (2018) \cite{Budish}, which considers a double-spend attack achieved through a hashrate rental market, and introduces a model where, in equilibrium, such an attack is relatively cheap. 
We first modify this model to show that, without considering counterattacks, double-spend attacks may be either free, cheap or
impossible, depending on the market availability of rental hashrate and the price impact
of an attack. We then extend Budish's model by also allowing  the victim to have access to the same  hashrate market as the attacker. This leads to the War of Attrition model and the Retaliation game.

A number of other papers have also taken inspiration from Budish. Among them, Auer (2019) \cite{Auer} analyzes payment security for PoW cryptocurrencies and, independently and contemporaneously, derives a safety condition similar to our Corollary \ref{no-double-spends}. Auer shows that, taking the long term view (at some point, the block reward will be low), PoW security will come largely from transaction fees, which are only high when congestion is high. Auer shows that fees suffer from a free-rider problem, suggesting that in the long-term the security on any PoW chain will  be low, giving a similar result to Carlsten et al. (2016) \cite{Instability}. In future work, it will be interesting to examine whether our theoretical analysis of counterattacks may mitigate this observation about the role of fees. 

Following Auer, a white paper provides a similar  safety condition, but without considering  hashrate market impact (Hasu et al., 2019)~\cite{Hasu}. These authors emphasize that miners typically have high upfront costs (e.g., purchasing mining hardware and prepaying electricity costs), and that an attack that reduces the price of the mined asset will reduce the value of their mining hardware. Thus, miners are incentivized not to lease hashpower to potential attackers, making an attack more difficult.

The notion that a victim might be able to launch a counterattack has been raised before, but  largely dismissed as difficult to implement and ineffective. As we show in our model, the mere ability of recipients or some service they employ to counterattack may be sufficient to discourage attacks from happening in the first place.
Bonneau (2016) \cite{Bonneau} briefly discusses the possibility of counterattacks, but suggests that this would place undue burden on the recipient of a transaction to monitor the chain and be willing to counterattack.\footnote{As we discuss in Section \ref{Discussion}, we agree that the requirement to monitor the chain and be ready to counterattack is, at present, a somewhat difficult task and that this  changes the security model of PoW systems. However, it seems reasonable to study a system with this capability in place, especially given that our model shows that attacking is unprofitable when the victim is able to counterattack.}
Bonneau situates this discussion in the context of an attacker who can rent hashrate by bribing miners, raising the concern that the net cost of a double-spend attack can be low. Judmayer et al.\ (2019) \cite{Pay-To-Win}  also briefly consider the possibility of counterattacks, saying simply that if a defender counterattacks that this results in a ``bidding game''. They also show  how to efficiently implement various bribing-style attacks on Bitcoin, by guaranteeing  payments through smart-contracts on Ethereum. Lastly, Vorick (2019)~\cite{Vorick} informally analyzes a   game with retaliation, but without the assumption of decreasing profit from attack over time, and concludes that counterattacks would not deter attackers. 

Without considering counterattacks, Liao and Katz (2017)~\cite{Whale} analyze a specific form of Bonneau's bribing attack in which an attacker, lacking a majority of hashpower, incentivizes other miners to mine on the shorter attack chain by issuing large-fee transactions on that chain. Other miners can collect the large fees only if they join the attacker in mining on the attacking chain. Thus the attacking chain can gain majority support, making the 51\% attack successful, and the attacker only has to pay if the attack is successful. Our model considers that the attacker rents hashrate from a marketplace, but it can be abstracted to be equivalent to this model. Thus our retaliation argument applies to this attack as well.

There have also been  attempts to compare the economic security (that is, the difficulty of double-spending) on different PoW chains. A well-known but sometimes unreliable source for this information is Crypto51.app \cite{crypto51}, which reports the hashing algorithm and hashrate of each cryptocurrency. Using values from the NiceHash  hashrate marketplace~\cite{NiceHash}, it also estimates the cost of renting the equivalent of one hour's worth of that network's PoW, as well as the availability of that amount of hashrate in the marketplace. In a blog post, Carter (2019) \cite{Carter-Settlement} also ranks the security of cryptocurrencies in order of how much is paid to miners per unit time.

Broadening out from double-spending attacks, while still appealing  to decision-theoretic
and game-theoretic models, Eyal and Gun Sirer (2013) \cite{Selfish} consider a  miner who deviates from the default protocol in order to earn more block rewards.
In this model of {\em selfish mining}, they show that a miner with a significant but non-majority fraction of the mining power can earn more block rewards in expectation by  mining in secret and following a carefully designed, alternate mining policy. This  has in turn inspired a series of follow-up work. Among these, Sapirshtein et al.\ (2015) \cite{Opt-Selfish} encapsulates some features of the mining game into a  Markov decision process, and solves for the optimal selfish strategy using value iteration.
Related to selfish mining, Arnosti and Weinberg (2018) \cite{Oligopoly} and Leonardos et al.\ (2019) \cite{Oceanic} provide theoretical analyses that indicate miners may profit from concentrating their power, suggesting that PoW systems will grow increasingly centralized.
Lastly, Fiat et al.\ (2019) \cite{Energy-Equilibria} and Goran and Spiegelman (2019) \cite{Mind-Mining}  propose and analyze another profitable deviation from the honest PoW protocol: occasionally turning mining equipment off in order to reduce the difficulty threshold, making mining more profitable later. 

\subsubsection{Standard Game-Theoretic Models}

The double-spending retaliation model  that we introduce is related to two well-known game-theoretic models: the War of Attrition and the Volunteer's Dilemma. 

The War of Attrition, first analyzed by Maynard Smith (1974) \cite{Maynard-Smith-74}, models two animals fighting for a single resource. The longer the fight goes on, the worse off they both become. 
The unique symmetric equilibrium is a mixed equilibrium in which each player in each round fights with  decreasing probability.
This equilibrium is both an {\em evolutionarily stable strategy} (ESS) and a {\em subgame perfect equilibrium} (SPE), both of which are refinements of a Nash equilibrium. Game-theoretic models in the biology literature~\cite{Maynard-Smith-74} more typically use the concept of ESS, while analyses in the economics literature~\cite{Fudenberg-Tirole-Book} (Section 4.5.2) more typically use the concept of SPE. If there is some asymmetry between the players, for example if one player is in possession of the resource to begin with, Maynard Smith also shows that no-fighting can  be an ESS. 

The Volunteer's Dilemma, introduced in Diekmann (1985) \cite{VOD}, models a multi-player game in which one player must volunteer to pay a cost in order to provide all players with a benefit. 
Weesie (1993) \cite{Weesie} shows that if the value of the benefit is decreasing over time and there is an asymmetry in costs among the players, there is a unique SPE in which the player with the lowest cost volunteers immediately.
In our analysis, we take Weesie's model of decreasing value over time and asymmetric players and couple this with the War of Attrition model, achieving a  SPE for PoW security 
in which there is no attack in equilibrium.

\section{A Model of Majority Attack Without Retaliation}
\label{Budish-Section}

We first modify the model and analysis of a double-spend attack from Budish \cite{Budish}. As with Budish, we will assume the availability of a hashrate marketplace, but also allow for some friction, so that the rental cost increases as more hashpower is rented. We also allow for an attack having the effect of changing  the price of the cryptocurrency. 

Based on this, we determine the cost of a single double-spend attack in terms of the block reward. For now, we assume that the victim of the double-spend will not retaliate. The analysis reveals a {\em safety condition}: the size of a transaction must be small enough relative to the size of the block reward in order to make it unprofitable to double-spend the transaction.

\subsection{Model}

Consider the following elements of our model:
\begin{itemize}
\item Let $p_b>0$ denote the {\em block reward},\footnote{The block reward $p_b$ includes both the block subsidy and transaction fees paid to miners. For simplicity, we assume that $p_b$ is constant over the time we consider, even though transaction fees may vary. This is a reasonable approximation since, as of Febuary 2020, the typical total fee reward for a Bitcoin block is about 0.5\% - 2.5\% of the block subsidy \cite{fees}, so $p_b$ does not vary much block-to-block.} measured in dollars. 
\item Let $n>0$ denote the amount of honest hashpower in the system, measured in hashes per block.
\item Let $c_h>0$ denote the marginal cost of renting one unit of hashpower in the absence of attack, in dollars per hash. 
\item Let $\b \geq 0$ denote the multiple of the total honest hashpower that the attacker acquires. 
\item Let $e$ denote the escrow period, in blocks on the honest chain, that the victim waits before accepting
a transaction as valid.
\item Let $v>0$ denote the size of the transaction being attacked, in dollars.
\end{itemize}

We  allow for two kinds of friction resulting from an attack. 
The first kind of friction we consider is that the hashrate market may not be perfectly liquid, and the cost of renting one unit of hashpower may increase as an attacker rents hashpower. For this, let $\k(\b) \geq 0$ denote the market impact on the hashrate market as a function of $\b$, the fraction of honest hashrate that is rented. We assume that the function $\k(\b)$ is weakly monotone increasing as $\b$ increases.
For example, the cost of renting one unit of hashrate, meaning $\b=1/n$, is $(1+\k(1/n))c_h$.

The second kind of friction we consider is that the attack may have an effect on the price of the underlying cryptocurrency. For this, we use $0 \leq \D \leq 1$ to denote the price decrease, relative to dollars, of the underlying cryptocurrency after an attack, so that the price of one dollar's worth of that cryptocurrency is, after attack, reduced to be worth $(1-\D)$ dollars.\footnote{Our analysis of a single double-spend attack would also go through for a price increase (and negative $\D$),  just as long as the  hashrate market impact $\k(\b)$, satisfies $\k(\b)>|\D|$, so that the net cost of attack  remains positive.} Later, we will use this same approach to model the ongoing, additional effect on price of a sequence of attacks and counterattacks. 

\subsection{Analysis}

Suppose that anyone with hashpower is free to enter into the mining game of a permissionless network such as Bitcoin. Based on this, competition between miners for block rewards leads to an {\em equilibrium condition}, relating the amount of hashpower  in a PoW system to the mining reward.
\begin{lemma}[Free entry, Budish \cite{Budish}]
\label{simplelem}
Free entry into the mining competition and liquid hashpower leads to the equilibrium condition 
\begin{equation}
n = p_b / c_h. 
\end{equation}
\end{lemma}
\begin{proof}
Suppose that $n < p_b/ c_h$, meaning there exists an $\eps > 0$ for which $n + \eps < p_b / c_h$. 
In expectation a miner with $\eps$ hashpower who enters the mining competition will earn $\eps p_b/(n+\eps)$ but pay $\eps c_h$. By assumption $p_b/(n+\eps) > c_h$ and the miner would be better off entering. Thus this is not an equilibrium. A similar argument shows that the marginal miner whose payments  (in expectation) are greater than earnings will drop out. 
\end{proof}

Lemma~\ref{simplelem} provides a simple relationship between the amount of honest hashpower $n$, the block reward $p_b$, and the no attack, marginal cost of renting hashpower $c_h$.

Based on this, Theorem~\ref{cost_of_attack} derives the net cost of a single attack as a function of the block reward, $p_b$, the escrow period, $e$, the market impact  $\k(\b) \geq 0$ of a hashrate purchase of size $\b>1$ (majority is needed for attacks that we consider), and the price decrease of the underlying cryptocurrency after an attack, $0 \leq \D \leq 1$.

Theorem \ref{cost_of_attack} modifies an earlier analysis presented in Budish~\cite{Budish}.\footnote{We thank Eric Budish for helpful discussions regarding this formulation.}  
Budish models the net cost of attack, and establishes that the cost of attack can be quite small. In this earlier model, the attacker rents majority $\beta$ hashpower for $e$ honest-block-times but only collects $e$ block rewards, whereas in our analysis the attacker collects  $\beta e$ block rewards, reflecting the relative length of the attack chain.\footnote{We thank Jacob Leshno for pointing this out.} 
Without further modification, this analysis reveals a (likely unrealistic) zero net cost of attack (see Corollary \ref{zero_cost_corollary}). By considering one or both of the two different kinds of market frictions, we reintroduce a positive net cost of attack. As discussed above, one possibility is that the attacker's cost of hashrate is larger than the honest miners’ costs, represented by $\kappa(\beta)$. A second possibility is that the value of the underlying cryptocurrency declines as a result of the attack, represented by $\Delta$, which is a possibility also discussed by Budish.
\begin{theorem}[Net Cost of Attack]
\label{cost_of_attack}
The net cost of a single double-spend attack on a proof-of-work system, considering the possibility of hashrate market impact and currency value decrease, is 
\begin{equation}
(\k(\b) + \D) e p_b
\end{equation}
\end{theorem}
\begin{proof}
The cost of renting a $\b$ majority of mining power to generate $e$ blocks is $(1+\k(\b)) \b n c_h$ per honest-block-time, but only requires mining for $e/\b$ honest-block-times, giving a total cost of $(1+\k(\b))\b n c_h e/\b = (1+\k(\b)) e n c_h$. 
Since an attacker earns block rewards on the attacking chain, but these block rewards are reduced in value by $\D$, a successful attack earns $(1-\D) e p_b$. 
Substituting (1), the net cost of attack is $(1+\k(\b)) e p_b - (1-\D) e p_b = (\k(\b) + \D) e p_b$.
\end{proof}

Knowing the cost of a single double-spend attack, we can now derive a safety condition, sufficient to protect against double-spend attacks being profitable, as a function of the parameters in the above theorem and the size of the transaction being attacked, $v$.
\begin{corollary}[Safety condition]
\label{no-double-spends}
It is not profitable to conduct a single double-spend attack on a proof-of-work system, considering the possibility of hashrate market impact and currency value decrease, if and only if 

\begin{equation}
p_b > \frac{v}{(\k(\b) + \D)e}
\end{equation}
\end{corollary}

\begin{proof}
In order for no double-spend attack to be profitable, we need  the net cost of attack to be larger than the largest possible benefit from attack, i.e.,
$(\k(\b) + \D) e p_b > v$. 
\end{proof}

The way to think about Corollary~\ref{no-double-spends} is that if large transactions are to be safely supported, then $p_b$ needs to be correspondingly large such that if the attacker were to rent the amount of hashrate needed to execute the attack, the attacker would have to pay more than the stolen transaction would be worth. The large block reward thus incentivizes miners not to participate in a double-spend but to mine honestly instead (and at a greater profit). The network becomes safer as $p_b$ increases, and as $\k(\b), \D$, and $e$ increase. The escrow period $e$ is under the control of the potential victim.

As a special case of Theorem~\ref{cost_of_attack}, when $\k(\b)=0$ for all $\b$, that is the hashrate marketplace is liquid enough that the attacker's  purchase does not increase the hashrate price, and $\D=0$, that is the value of the cryptocurrency does not change on news of the attack, the cost of the attack is zero, and  no size of block reward or length of escrow period can disincentivize attack.
\begin{corollary}[Zero-Cost Attack]
\label{zero_cost_corollary}
if $\D=0$ and there exists $\b>1$ such that $\k(\b)=0$, then the net cost of a single double-spend attack is zero, and it is always profitable to attack a transaction of non-zero value.
\end{corollary}
\begin{proof}
Substitute $\k(\b)=0$ and $\D=0$ into Theorem~\ref{cost_of_attack}. 
\end{proof}

This corollary implies that, for blockchains for which the hashrate marketplace is very liquid relative to the total amount of work done on the chain, and for which the price does not change upon news of attack, \emph{every} transaction is susceptible to a  double-spend attack no matter the escrow period. 

\section{The Retaliation Game}
\label{Our-Model}

Going forward, we make the assumption that at least one of  $\Delta$ or $\kappa(\b)$ (for all $\b>1$) is non-zero, such that we have a (possibly small) positive cost of attack, and we denote this {\em  cost of attack} as $c>0$. 
Our main result is to  establish that the possibility of retaliation can amplify this cost of attack until it is no longer profitable in any circumstance (e.g., under any valid $v$, $e$ and $p_b$). 

In the Retaliation game, we allow a victim who has been double-spent to counterattack. Just as the attacker might rent sufficient hashpower to conduct a double-spend from a hashrate marketplace, a victim might rent from the same marketplace at the same cost to retrieve its property. In this way, the interaction between the attacker and the victim takes the shape of a War of Attrition, in which two players compete for a fixed prize by opting to attack each other in turns, where each time they attack this incurs an additional cost. 

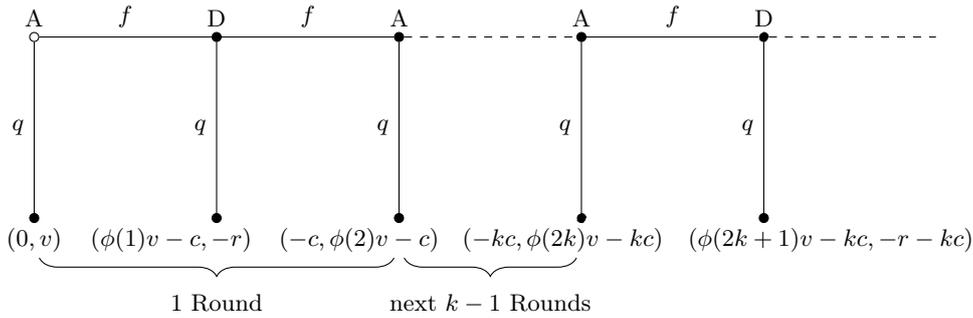
\begin{figure*}[!t] 
\centering
\begin{tikzpicture}[font=\footnotesize,scale=1.6]
\tikzstyle{solid node}=[circle,draw,inner sep=1.2,fill=black];
\tikzstyle{hollow node}=[circle,draw,inner sep=1.2];

\node(0)[hollow node]{}
child[grow=down]{node[solid node]{}
	edge from parent node[left]{$q$}}
child[grow=right]{node(1)[solid node]{}
child[grow=down]{node[solid node]{}
	edge from parent node[left]{$q$}}
child[grow=right]{node(2)[solid node]{}
child[grow=down]{node[solid node]{}
	edge from parent node[left]{$q$}}
child[grow=right]{node(3)[solid node]{}
child[grow=down]{node[solid node]{}
	edge from parent[solid] node[left]{$q$}}
child[grow=right]{node(4)[solid node]{}
child[grow=down]{node[solid node]{}
	edge from parent[solid] node[left]{$q$}}
child[grow=right]{node(5){}
edge from parent[dashed] node[above]{}}
edge from parent[solid] node[above]{$f$}}
edge from parent[dashed] node[above]{}}
edge from parent node[above]{$f$}}
edge from parent node[above]{$f$}};

\foreach \x in {0,2,3}
\node[above]at(\x){A};
\foreach \x in {1,4}
\node[above]at(\x){D};

\node[below]at(0-1){($0, v$)};
\node[below, xshift=-17]at(1-1){($\p(1)v-c,-r$)};
\node[below, xshift=-15]at(2-1){($-c,\p(2)v-c$)};
\node[below, xshift=-7]at(3-1){($-kc,\p(2k)v-kc$)};
\node[below, xshift=25]at(4-1){($\p(2k+1)v-kc,-r-kc$)};

\draw[decoration={brace,mirror,raise=15pt,amplitude=7pt},decorate]
(0-1) -- node[below=25pt] {1 Round} (2-1);
\draw[decoration={brace,mirror,raise=15pt,amplitude=7pt},decorate]
(2-1) -- node[below=25pt] {next $k-1$ Rounds} (3-1);

\end{tikzpicture}
\caption{An extensive-form description of the Retaliation game.
The (A)ttacker has the option to begin the game by fighting, stealing a transaction worth $v$ from the (D)efender. $D$ has the option to counterattack. Each attack, by either $A$ or $D$, costs $c>0$. Decreasing function $\p(t)$ models the price decrease of the transaction  $v$ as a function of the number of  attacks completed $t$ (starting with $t=0$), and $r>0$ models the value of $D$'s reputation loss if $D$ loses. Since each step $t$ in this sequence counts a ``half-round,'' the round number $k=\lfloor t / 2 \rfloor +1$. The game ends when either player quits. Each terminal node lists the payoffs to ($A$, $D$). The payoff depends on the round, and on which player quits. \label{woa-2}}
\end{figure*}

Our results crucially rely on the net profit of each successive attack (value of cryptocurrency gained minus cost of attack) falling for at least one of the players, so that it eventually becomes unprofitable to counterattack one more time. This ensures an end to the game because at that point it will be more damaging for a player to counterattack than to quit. We think this is a mild assumption, as it can be achieved in a number of ways.  For example,  it can arise by assuming that each attack weakens the value of the underlying cryptocurrency,  while the cost of attack remains relatively constant. Decreasing profit could also arise if the value of the cryptocurrency stays relatively constant but the cost of attacking increases over time (e.g. because a player's cost of capital rises as the player borrows more capital to finance the escalating counterattacks). 

\subsection{The Multi-Round Double-Spending Model}

See Figure \ref{woa-2} for an extensive-form description of this multi-round, double-spending game with retaliation (the {\em Retaliation game}). 

We need a small amount of new notation. 
We consider two players, an (A)ttacker and a (D)efender.
As explained above, we adopt $c>0$ to denote the net cost, in dollars, of conducting a single, double-spend attack.
This is  the net cost of an attack on the network by the attacker
as well as the net cost of an attack on the network by the victim, who can choose to launch a retaliation attack.\footnote{The initial attack, due to the escrow period, typically will require a larger reorg than each successive attack, which merely needs to surpass the previously-overtaking chain. For simplicity, we keep $c$ constant throughout, though this asymmetry is an additional, initial advantage to the defender.}
We continue to use  $v>0$ to denote the value, in dollars, of the transaction that is in conflict.
Attacker $A$ has the initial option to attack. Thereafter, players alternate choosing whether to (f)ight or (q)uit, with q ending the game and leaving the opponent with the reward. The game can, in principle, continue indefinitely. We refer to each moment at which one of the two players 
has the option to fight or quit as a {\em time period}, with $A$ making the first decision at time $t=0$. 
We refer to a {\em round} as a play of $A$ followed by a play of $D$.\footnote{The game could be made more complete by including strategic miners who take a loss when reorgs are revealed and their mining rewards taken from them. This could provide an additional initial asymmetry in favor of the defender, but for simplicity we consider that all miners besides $A$ and $D$ follow the heaviest chain rule.}
 
We assume the net profit from each successive attack falls until at some point it becomes negative for at least one of the players. For concreteness, we do this through the following two specific assumptions (but as discussed above, there are other reasons  why the net cost of attack can fall):
\begin{enumerate}
\item The value of the underlying cryptocurrency decreases with the number of  attacks (including counterattacks), making the reward worth $\p(t) v$ after $t$ total completed attacks, where $A$ makes the first move at $t=0$, $\p(0)=1$, and $\p$ is a function that is monotonically decreasing in the number of time periods $t$. 
For example, we could have $\p(t)=(1-\D)^t$, for $0 < \D < 1$, referring to the earlier use of $\D$ to model the multiplicative price decrease of the cryptocurrency after a successful attack. Here, $\D = 0.1$ would mean the price decreases by 10\% of its contemporaneous value after each attack. Alternatively, we could have $\p(t)=\max(1-\g t,0)$ for $0 < \g \leq 1$, which would take value $0$ for all $t \geq 1/\g$. Here, $\g=0.1$ would mean the price decreases by 10\% of the pre-attack value after each successive attack.
\item The net cost of attack,
$c > 0$, remains constant even as the underlying price of the cryptocurrency falls. Note it is not crucial that that $c$ remains constant. If $\p(t)$ is decreasing, $c$ merely needs to decrease more slowly than $\p(t)$ (or $c$ could increase); we assume it remains constant for notational convenience. 
\end{enumerate}
 
In addition,  we assume that quitting is free for $A$,  but costs {\em reputation cost} $r>0$ for $D$. To motivate this, we imagine a typical defender to be an exchange with a public reputation to lose, whereas a typical attacker is  anonymous. If an exchange is attacked, customers may be less likely to use the exchange in the future. This introduces an asymmetry into the model.
We discuss the assumptions of the model in more detail in Section \ref{Assumptions}.

\subsection{Equilibrium Analysis}

The strategy of a player defines, for each possible history at which it has an action to choose, the probability $p$ with which to fight. 
A complete strategy profile can be written as $\s = \{p_0, p_1, p_2, \ldots p_t, \ldots\}$, with $p_t$ the probability of fight by $A$ if time period $t$ is even, and the probability of fight by $D$ otherwise. Since each step $t$ in this sequence counts a ``half-round'', the round number $k=\lfloor t / 2 \rfloor +1$.

We define the {\em break-even time} $T_{\{A,D\}}$, for each player $A$ and $D$, as the first point in time at which quitting is weakly more profitable than continuing.\footnote{The analysis follows some aspects of the analysis of the Volunteers Dilemma \cite{Weesie}. In both cases, we have decreasing $\p$, a benefit, and a cost. The principal difference is that in our model just having one actor pay the cost does not make the benefit free for the other actor. This property makes our game similar to the War of Attrition, in which both players pay the cost of escalating the game.} These break-even times are continuous quantities, and delineate distinct ranges of time steps that play a role in the theoretical analysis. 

For the attacker, $T_A$ satisfies 
\begin{align}\p(T_A)v-c &= 0,
\end{align}
so we have $T_A = \p^{-1}(c/v)$. 
For the defender, $T_D$ satisfies 
\begin{align}
    \p(T_D)v-c = -r,
    \end{align}
    so we have 
$T_D = \p^{-1}((c-r)/v)$. 

Due to our restrictions on $\p$ to be non-negative, if $r>c$ then $T_D$ is undefined, and it is always profitable for $D$ to fight. 
Since $T_A$ is guaranteed to be finite, $A$ will surely quit at $\lceil T_A \rceil$, and there is no chance of an infinite game.

Suppose without loss of generality that player $i\in \{A,D\}$ has  $T_i\leq T_j$, where $j$ denotes the other player. Now we recognize that player $i$ only needs to consider time steps with $t \leq \floor{T_i}$.  If $t > \floor{T_i}$, player $i$ is sure to choose quit, and  the game will end. Opponent $j$ similarly needs only to consider turns in which the time step $t$ satisfies $t \leq \floor{T_i} + 1$. For this reason, it is without loss of generality to consider a finite strategy profile, defining all actions until one step beyond $T^* = \floor{\min\{T_A, T_D\}}$. Beyond this point, quit is a dominant strategy for at least one of the two players, ending the game surely.

We now provide the main theoretical result, which states that a non-fighting strategy profile is a SPE of the Retaliation game. 
The proof makes use of the one-deviation principle, i.e., 
that a strategy profile is a SPE in a finite, extensive-form game, if and only if there is no profitable single deviation. Our game has an equivalent, finite form, because of time  $T^*$, beyond which quit is a dominant strategy for at least one of the two players, which ends the game at that time.

In the following, it will be convenient to let  $\odd{t}$ denote the largest odd integer less than or equal to $t$, and  $\even{t}$ denote the largest even integer less than or equal to $t$. In particular, $\odd{T_D}$ denotes the last time step at which $D$ has a profitable move and $\even{T_A}$ denotes the last time step at which $A$ has a profitable move.
\begin{theorem}[No-attack equilibrium]
\label{no-attack-equilibrium}
In the Retaliation game, if the defender $D$ is last to have a profitable move, then the following strategy profile is a subgame perfect equilibrium: 
\begin{align}
    p_t&=\left\{
    \begin{array}{ll}
    1 & \mbox{if it is $D$'s turn, and $t\leq T_D$ (i.e., the attack is still profitable for $D$),}\\
    0 & \mbox{otherwise.}
    \end{array}
    \right.
\end{align}
\end{theorem}
\begin{proof}
Recall in the following that  $A$ and $D$ play on even and odd turns, respectively. Formally, $D$ is last to have a profitable move if $\even{T_A}<\odd{T_D}$. 

We proceed by case analysis:

(Odd time step $t \leq \odd{T_D}$, $D$ to play). Following the SPE, $D$'s utility will be $\p(t)v-c \geq -r$ 
since the SPE necessitates $p_{t+1}=1$. If $D$ deviates and plays $p_t<1$, the expected utility will be $p_t(\p(t)v-c) + (1-p_t)(-r) \leq \p(t)v-c$ by assumption that $t \leq T_D$ and the definition of $T_D$. Therefore $D$ has no incentive to deviate. 

(Odd time step $t > \odd{T_D}$, $D$ to play). Similarly, by definition of $T_D$, we see that $p_t>0$ reduces utility compared with $p_t=0$.

(Even time step $t > \even{T_A}$, $A$ to play). Deviating from the SPE and selecting $p_t>0$ leads to negative utility in every case while $p_t=0$ leads to 0 utility. 

(Even time step $t \leq \even{T_A}$, $A$ to play). Since we assumed $\even{T_A}<\odd{T_D}$, $t+1 \leq \odd{T_D}$, thus $p_{t+1}=1$ by the SPE. Given this, $A$ has no chance of winning in this round and therefore $A$'s utility will be $-p_t c$ if $p_t>0$, with utility $0$ if $p_t=0$. No deviation is profitable here.

\end{proof}

As long as $D$ is the last to have a profitable move, then the equilibrium is for $A$ to play `never fight' ($p_t=0$) and $D$ to play `always fight' ($p_t=1$). The effect of this is a no-attack equilibrium, because $A$ does not attack when given the chance, and $D$ does not need to counterattack. 
The main idea is that after a point in time, $\odd{T_A}$, it will no longer be in $A$'s  interest to fight because the cost of an attack will be higher than the best possible benefit, while quitting will net zero utility. Meanwhile, $D$ has the last profitable move, and is still willing to fight at least in the next turn after $\odd{T_A}$. Based on this, in every turn in which it may be profitable for the attacker to fight, the defender is sure to fight back, making it best for the attacker to immediately quit.

We provide a  sufficient condition for $D$  to be last to have a profitable move, and for this no-attack equilibrium. From the asymmetry provided by reputation cost $r>0$, we know that $D$ has a worse quitting outcome than $A$, and that $T_A<T_D$, for hold-out times $T_D$ and $T_A$. The hold-out times are continuous, and we need the difference to be large enough to imply  $\even{T_A}<\odd{T_D}$. 
We establish a safety condition on reputation cost $r$ for the special case of a linear effect of attack on the value of the cryptocurrency.

\begin{theorem}[Reputation safety condition]
\label{rep-safety-condition}
In the Retaliation game, given $\p(t)$ of the form $\p(t)=\max(1-\g t,0)$, with $0 < \g \leq 1$,  then
\begin{align}
r > \g v
\end{align}
is sufficient for the no-attack equilibrium.  
\end{theorem}
\begin{proof}
Concretely, $D$ having the last profitable move is equal to condition $\odd{T_D} > \even{T_A}$. This is always satisfied when $T_D > T_A + 1$, since applying the odd function to both sides of this inequality gives us $\odd{T_D} \geq \odd{T_A + 1}$. It can be shown 
that $\odd{T_A + 1} = \even{T_A}+1 > \even{T_A}$, thus showing $\odd{T_D} > \even{T_A}$. 

We now derive an equivalent condition for $T_D > T_A + 1$ involving $r$. Substituting the definitions of $T_D$ and $T_A$, we seek $r$ such that $\p^{-1}(\frac{c-r}{v}) > \p^{-1}(\frac{c}{v}) + 1$. Applying the monotonically decreasing function $\p$ to both sides of the inequality, and thus reversing the inequality, we have $\frac{c-r}{v} < \p(\p^{-1}(\frac{c}{v}) + 1)$. Rearranging, this leads to
\begin{align}
r > c - v \p \left( \p^{-1} \left(\frac{c}{v} \right) + 1 \right).
\label{eq:r_condition_general}
\end{align}

We have $v>0$ and $c>0$. We have $c<v$ (otherwise the game would never start). We have that $\p^{-1}(t)$ is defined and non-negative for positive arguments $0<t\leq 1$, and $\p$ is positive for non-negative arguments, so $v \p ( \p^{-1} (\frac{c}{v} ) + 1 ) > 0$. 

Given $\p(t) = \max(1-\g t,0)$, $t\geq 0$ with $0<\g<1$, we have $\p^{-1}(t) = \frac{1}{\g}- \frac{1}{\g}t, 0< t \leq 1$. Note we only consider these two functions on their valid domains in the following argument.
\begin{lemma}
Given $\p(t) = \max(1-\g t,0)$ with $0<\g<1$, $\p^{-1}(\frac{c}{v} - \g) \geq \p^{-1}(\frac{c}{v})+1$
\label{lemma:phi_inv_size}
\end{lemma}
\begin{proof}
Suppose for the purpose of contradiction we have $\p^{-1}(\frac{c}{v} - \g) < \p^{-1}(\frac{c}{v})+1$. Applying the definition of $\p^{-1}(t)$ to the inequality, we have $\frac{1}{\g} - \frac{1}{\g}(\frac{c}{v} - \g) < \frac{1}{\g} - \frac{1}{\g}(\frac{c}{v}) + 1$. Simplifying, we have $ -\frac{c}{\g v} + 1 < -\frac{c}{\g v} + 1$. This is a contradiction.
\end{proof}

\smallskip

Since $\p$ is decreasing, we can now substitute the result of Lemma \ref{lemma:phi_inv_size} into~\eqref{eq:r_condition_general}, to get 
$r > c - v \p ( \p^{-1} (\frac{c}{v}) + 1) >c - v \p ( \p^{-1}(\frac{c}{v} - \g))
=  c-v(\frac{c}{v}-\g) = \g v$.\sloppy

\end{proof}

\smallskip

As an example, if $\g=0.1$ and $v=10000$, $D$ is sure to win if $D$'s loss of reputation from being successfully double-spent is greater than $r = 1000$.

\subsection{Modeling Assumptions}
\label{Assumptions}

In this section we will further discuss the specific modeling assumptions we make.

\subsubsection{The net profit of each successive attack falls}
The net profit of an attack has two parts: net cost of attack, that is the price of renting sufficient hashrate minus the mining rewards earned, and benefit from attack, that is the value of the transaction stolen. For concreteness, we have achieved this assumption by assuming that (1) the price of the asset weakens as attacks increase, and (2) that cost of attack stays constant. However, the model would work equally well assuming, for example, that the price remains constant and the cost of attack increases over time. Here we discuss some explanations for these possible assumptions.

\textit{Every double-spend attack further weakens the value of the underlying asset.}
This is a widely-held assumption, and one formally incorporated into at least one of the models put forward in each of Budish (2018) \cite{Budish}, Auer (2019) \cite{Auer}, and Hasu et al.\ (2019) \cite{Hasu}.\footnote{An industry commentator has also stated, ``A successful 51\% attack is likely to have a very negative effect on the market value of a cryptocurrency''~\cite{Consensys-Centralization}.} 
That this assumption is so common is perhaps surprising given that, for the few double-spend attacks that have been observed (see Table \ref{table:known-double-spends}), the price has not always been significantly affected and has not always gone down. The attack on BTG in May 2018 was followed by a decline in price, while the attack on LCC in July 2019 was followed by a relatively stable price, and the attack on BTG in Jan 2020 was followed by a rise in price.

\textit{Constant net cost of attack across rounds.}
We assume that the net cost of attack, $c>0$, remains constant even as the underlying price of the cryptocurrency falls. In fact, the proof of Theorem~\ref{no-attack-equilibrium} only requires  that the net cost $c$ decreases more slowly than the decrease in the value of the cryptocurrency, $\p(t)$. This assumption can be justified by recognizing that an attack is likely to throw the mining hashrate market out of equilibrium, keeping price higher than its equilibrium value because mining operators have fixed costs, electricity contracts, and other frictions that can prevent them from acting quickly. 

\subsubsection{Defender has a higher penalty to losing than the Attacker.}
We need this assumption, which encapsulates an asymmetry between the attacker and victim. Businesses such as exchanges, which are typical targets for attack, have large footprints. Since double-spends become public information (see Table \ref{table:known-double-spends}) and the cryptocurrency addresses of prominent groups are often widely known, if the wallet of a known entity $D$ is double-spent, the public
would know about this. $D$'s future customers may be more likely to engage in double-spend attacks against $D$ as a result, and others may be less likely to bring their business to the exchange. For these reasons, it is reasonable to assume that $D$'s value of losing is $-r$, for some reputation cost $r>0$, while cybercriminals who mount double-spend attacks remain anonymous, presumably for fear of prosecution. 

\section{Empirical Investigation} 
\label{Empirical-Investigations}

No hashrate-driven double-spend attacks are known to have occurred against Bitcoin or Ethereum since 2015.\footnote{In 2013, a double-spend attack was successfully executed against Bitcoin. However, the fork was due to a bug in the software that caused the network to split and follow different forks, not because the attacker  acquired a majority of the hashpower~\cite{BIP50,double-spend-bitcointalk}}
This is despite the existence of many large transactions that would appear profitable to double-spend if enough hashrate could be acquired (see Figure \ref{fig:btc-txn-freq}, Figure \ref{fig:eth-txn-freq}, and Table \ref{table:big-btc-txns}). One reason for this might be that the transactions are between trusted parties (or are actually sent between two wallets owned by the same party) who have a relationship outside the blockchain.

Despite the lack of attacks, mining pool concentration has caused concern in the Bitcoin and Ethereum communities. As of February 2020, the total Bitcoin hashrate was about $102 \times 10^{18}$ SHA-256 hashes per second, and four mining pool operators controlled a majority of the hashrate on the Bitcoin network, while two pools controlled a majority of the  hashrate on the Ethereum network (see Table \ref{table:btc-eth-pools}) \cite{Bitcoin-Difficulty,Bitcoin-Pool-Centralization,Top-ETH-Miners}. One implication of this is that Bitcoin and Ethereum may themselves be vulnerable to majority attacks if mining pools collude. Indeed, when a hacker exploited a software security vulnerability to steal \$40M worth of Bitcoin from Binance (a large exchange) in May 2019, the Binance CEO considered (but ultimately abandoned) the idea of recruiting large mining pool operators to double-spend the stolen Bitcoin back to the company~\cite{CZ-Reorg}.

It should be noted, however, that the operators of mining pools do not always directly control the mining hardware within the pool. Individual miners could leave a pool upon noticing it is participating in an attack. Indeed, in 2014 a Bitcoin mining pool, Ghash.io, obtained more than 51\% of the total Bitcoin hashrate. In response, users threatened to leave the pool, leading Ghash.io to commit to never obtaining more than 40\% of the Bitcoin hashrate~\cite{ghash}. As emphasised in \cite{Hasu}, miners may have a long-term stake in a particular chain remaining attack-free. 

However, a second implication of the small number of pools is that smaller chains such as Bitcoin Cash (BCH) and Ethereum Classic (ETC) should be vulnerable to attack because they use the same PoW algorithms as chains with a much larger amount of hashrate. In fact, ETC was successfully attacked a dozen times in Jan 2019 \cite{ETC-Attacked}. 
As of February 2020, the Bitcoin Cash network computed about $4.9 \times 10^{18}$ hashes per second \cite{BCH-Difficulty}. 
This means that if one of the larger Bitcoin mining pools was to redirect its hashrate (about 16\% of the total, see Table \ref{table:btc-eth-pools}) towards mining Bitcoin Cash, as it could without significant modification due to a shared hash function, it would provide over triple the entire Bitcoin Cash hashrate. Indeed, several chains have been attacked. This is also likely due to the greater relative availability of their hashrate on marketplaces like Nicehash \cite{crypto51}.

\subsection{Observed Attacks}

\begin{figure}
\centering
\begin{minipage}{.5\textwidth}
  \centering
  \includegraphics[width=0.99\linewidth]{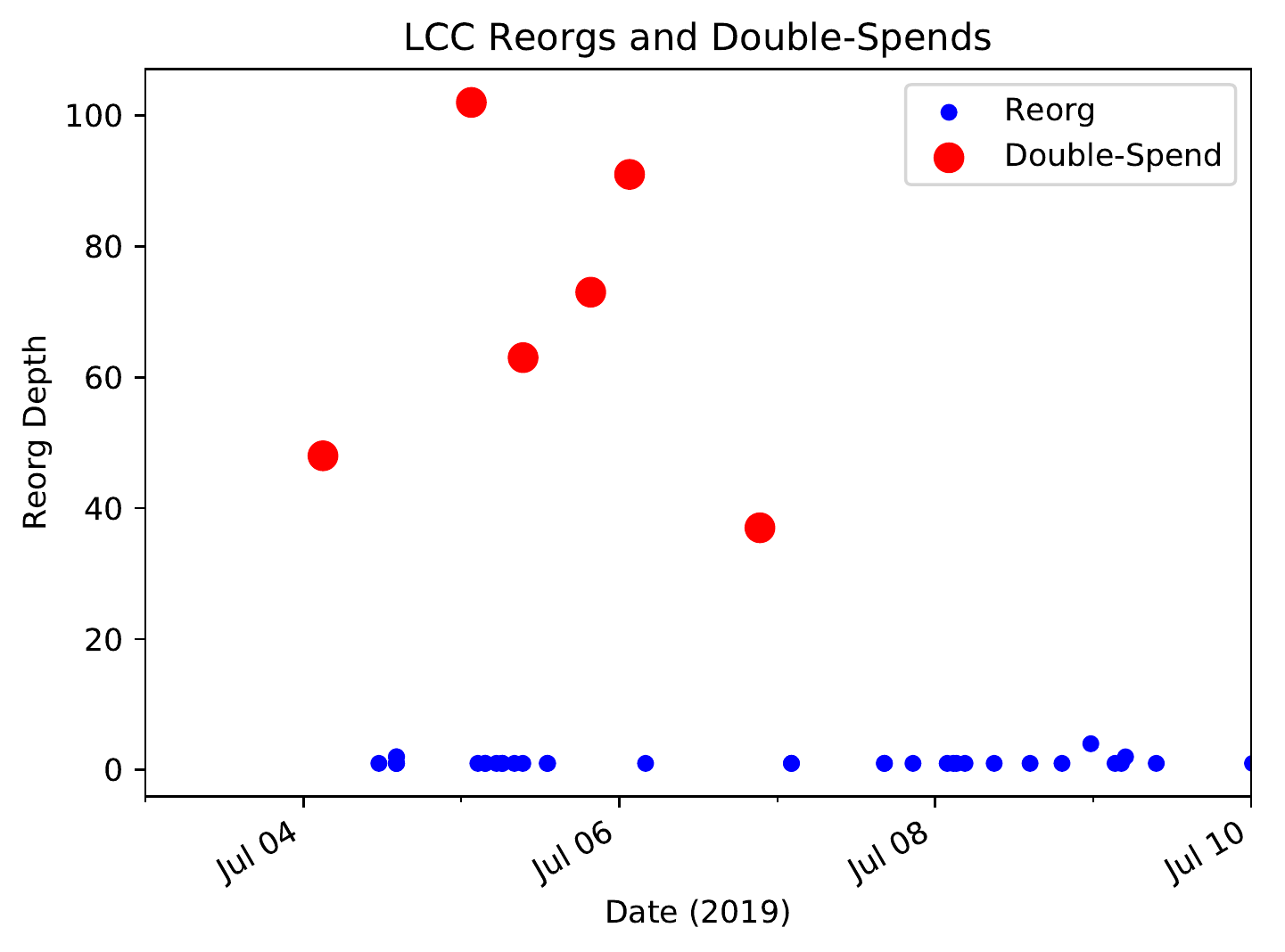}
  \caption{6 LCC Double-Spends, Jul-Sep 2019}
  \label{fig:lcc-doublespend-zoom}
\end{minipage}%
\begin{minipage}{.5\textwidth}
  \centering
  \includegraphics[width=0.97\linewidth]{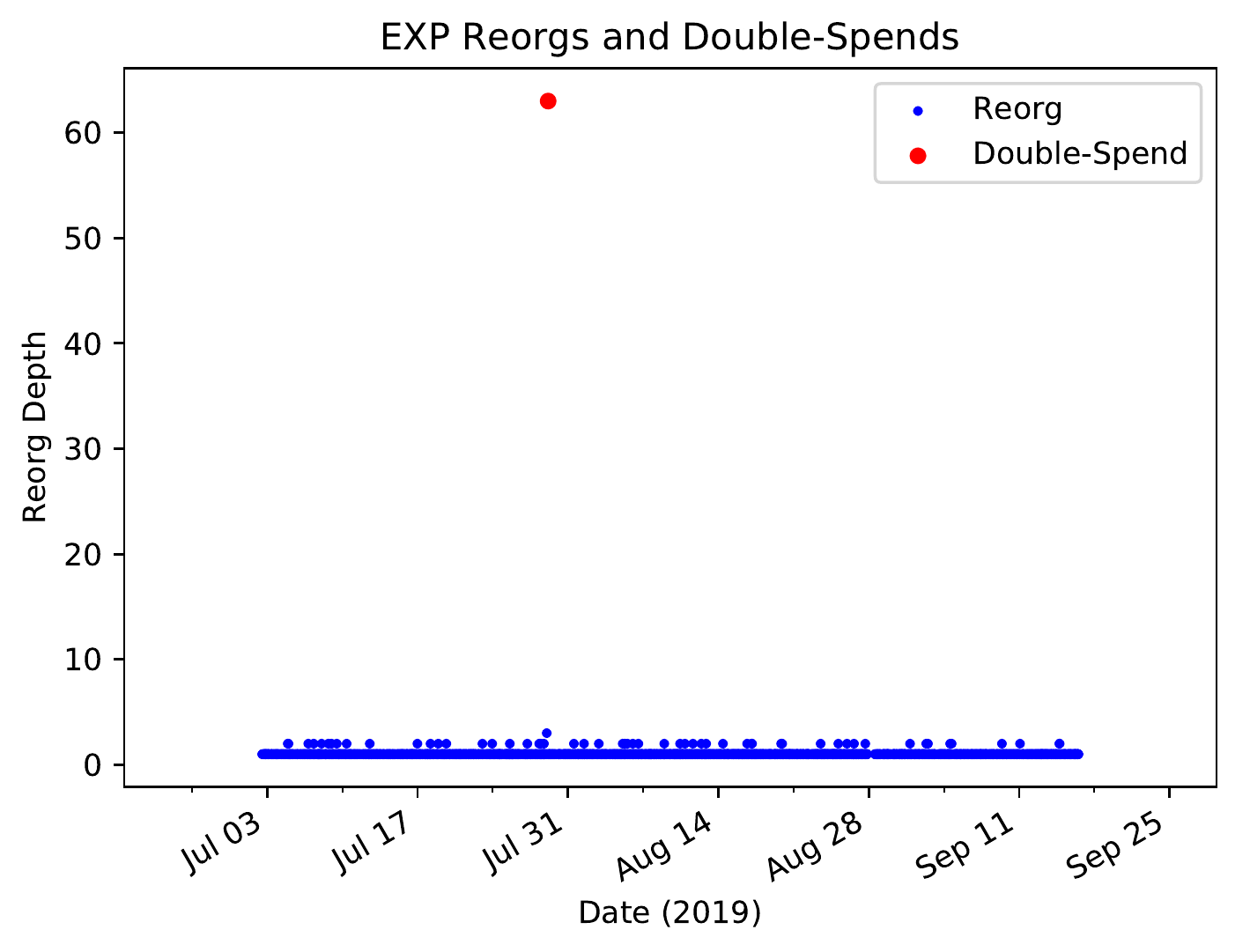}
  \caption{An EXP Double-Spend, Jul 2019}
  \label{fig:exp-doublespend}
\end{minipage}
\end{figure}

A list of  known double-spend attacks appear in Table~\ref{table:known-double-spends}, including attacks on the BTG, VTC, LCC and EXP chains that were recently discovered through the Reorg Tracker  system~\cite{BTG-Attacked}, a tool that logs reorgs and makes the data publicly available. The Reorg Tracker has, between July 2019 and February 2020, monitored 20 PoW chains
chosen based on their perceived susceptibility to hashrate rental attacks as well as their market capitalization.\footnote{BCH, BSV, BTC, BTG, CLO, DBIX, DOGE, ETC, ETH, EXP, IMG, LCC, LTC, MONA, PAC, PIRL, VTC, ZCL, ZEC, and ZEL} Most reorgs are non-malicious and do not contain any double spends (we call these random reorgs). They are typically naturally-occurring, low-depth (1 or 2 block) reorgs that happen when different miners find a block at the same height around the same time. The difference between random reorgs (shallow and frequent) and double-spends (deep and rare) is exemplified in Figure \ref{fig:lcc-doublespend-zoom} and Figure \ref{fig:exp-doublespend}.
The Reorg Tracker has observed a total of 18 double-spend attacks on four cryptocurrencies during this time, and as of Feb 2020, four of these attacks on Bitcoin Gold appear to be counterattacks. We now report on the double-spend attacks known publicly.

By far the largest known group of double-spend attacks (totaling over \$17 million) occurred on the Bitcoin Gold (BTG) network between May 16 2018 and May 19 2018 \cite{BTG-Verge-Attacked}. On May 18, the BTG Director of Communications advised exchanges to ``increase confirmations and carefully review large deposits'' \cite{btg_iskra}. One week later, on May 24, the BTG team put out a proposal \cite{btg_iskra_responding} to do a hard-fork (preserve user balances but update the blockchain software in a non-backward-compatible way) to a new hash function, in hope of preventing further double-spends by attackers who could accumulate hashpower of the old hash function. The update was successfully implemented but BTG was still not safe. 

In late January 2020, the Reorg Tracker observed two reorgs totalling about \$70,000 on Bitcoin Gold (BTG), each about 15 blocks deep. This was larger than the the contemporaneous 12-block BTG withdrawal requirement on Binance, which Binance increased to 20 blocks after the attack \cite{BTG-Attacked}. In early February 2020, the Reorg Tracker also observed preliminary evidence of retaliations to double-spend attacks on Bitcoin Gold.

As of mid February 2020, preliminary evidence from the Reorg Tracker showed eight reorgs totalling about \$120,000 on Bitcoin Gold (BTG), some of which appeared to be counterattacks. These appear to include three separate instances of a retaliation game, one of length four (attack, counterattack, and two further moves) and two of length two (attack and counterattack). The identities of the participants are not known.

Beyond BTG, six double-spend attacks were on Litecoin Cash (LCC)~\cite{LCC-Attacked}. These comprised a series of six deep reorgs between depth 40-100 over four days. These reorgs were far outside the norm of the frequent random reorgs which occur due to network latency issues and are depth 1 or 2, see Figure~\ref{fig:lcc-doublespend-zoom}.  All the original transactions were made to the same address and the double spend transactions redirected to a single different address, indicating one attacker and one victim. 

One double-spend attack observed by the Reorg Tracker was on the Expanse (EXP) network \cite{EXP-Attacked}. The victim account in this case is linked to a wallet with about 90\% of the total EXP supply, suggesting the victim is an exchange, likely Bittrex or Upbit, which host most of the EXP trading.

Figures \ref{fig:lcc-doublespend-zoom} and \ref{fig:exp-doublespend} depict the reorg-depth of double-spends and random reorgs as observed by the Reorg Tracker on Litecoin Cash (LCC) and Expanse (EXP), respectively. Red dots represent the double-spend attacks, with the y-axis recording the number of blocks removed from the chain during an attack. Blue dots represent naturally-occurring, low-depth (1 or 2 block) reorgs that happen when different miners find a block at the same height around the same time.

The total amount stolen by way of double-spend attacks is far less than the amount estimated to have been stolen by hacking methods involving thefts of private keys, or ransomware for example.  The Wall Street Journal, for example, reported that over \$1.5 billion USD in cryptocurrency has been stolen between 2014 and 2018 in over 56 prominent attacks~\cite{wsj-autonomous}. In contrast to double-spend attacks, stolen-key or ransomware attacks do not require subverting an existing financial transfer between the attacker and victim to succeed. We do not consider these types of attacks here because we seek to investigate the security of the underlying PoW system, not of the software stack it runs on. We are not aware of any other incentive attacks in the underlying PoW systems that enable theft of cryptocurrencies. 

\begin{table}[ht]
\centering
	\begin{tabular}{llllrr}
		\toprule
		Coin & Date(s) & Victim & Attacks & USD & Source \\ 
		\midrule
		BTG & 02.08.20 - 02.11.20 & Unknown & 8 & $120,000$ & RT \\
		BTG & 01.23.20 - 1.24.20 & Unknown & 2 & $70,000$ & RT \cite{BTG-Attacked}\\
		VTC & 12.01.19 & Unknown & 1 & $30$ & RT \cite{VTC-Attacked}\\
		EXP & 07.29.19 & Unknown & 1 & $10$ & RT \cite{EXP-Attacked}\\
		LCC  & 07.04.19 - 07.07.19 & Unknown & 6 & $50,000$ & RT \cite{LCC-Attacked}\\
		ETC & 01.05.19 - 01.08.19 & Bitrue & 12 & $1,100,000$ & \cite{ETC-Attacked}\\
		BTG  & 05.16.18 & Bittrex & ? & $17,500,000$ &  \cite{BTG-Verge-Attacked}\\
		VTC & 10.12.18 - 10.18.18 & Unknown & ? & $100,000$ & \cite{VTC-Attacked}\\
		Zencash & 06.02.18 & Unknown & ? & $600,000$ & \cite{Zencash-Attacked}\\
		\bottomrule
	\end{tabular}
	\caption{Double-Spend Attacks. The USD column is the USD equivalent size of the theft, in Millions of dollars. The notation `RT' next to the reference indicates that the source of our data is the Reorg Tracker \cite{BTG-Attacked}, which has monitored reorgs on 20 PoW chains from July 2019-Feb 2020.}
\label{table:known-double-spends}
\end{table}

\section{Discussion}
\label{Discussion}

\paragraph{Both double-spend attacks and counterattacks are technically difficult to implement.}
Perhaps a reason that we have seen relatively few double-spend attacks is that they are technically cumbersome to execute. The Nicehash marketplace has a difficult user interface to understand. Furthermore, conducting a double-spend likely requires custom mining pool software, as every existing mining pool is likely to be following the honest protocol. Counterattacking, as Bonneau \cite{Bonneau} wrote, is an even more challenging task because all of this technical work must be prepared ahead of time and set to activate automatically upon the detection of a deep reorg. This may explain why we have only begun to observed counterattacks in the empirical evidence. However, as the market matures (e.g. as new marketplaces like Honey Lemon \cite{honeylemon}, a cloud-mining market aggregator, come online) we expect these tasks to become easier. 

\paragraph{External trust enables parties to make large transactions without the risk of double-spending.}
It is possible that  transactions larger than the safety limit (Corollary \ref{no-double-spends}) only occur between parties who have  reasons to trust each other. As an example, most exchanges require their customers to disclose their legal identities and tax ID numbers, especially for significant deposits or withdrawals (see \cite{KYC} for an analysis of the information that exchanges collect on their customers). The threat of legal action by the victim (exchange) against the attacker (double-spending customer) likely incentivizes non-attack in this case. 

\paragraph{Market impact, $\kappa(\beta)$, is high for Bitcoin and Ethereum, protecting them from double spends, and low for other chains.}
As of Febuary 2020, NiceHash only has enough SHA-256 hashrate for sale (500 PH/s) to match about 0.5\% of Bitcoin's total SHA-256 hashrate (100,000 PH/s), and only had enough EthHash for sale (3TH/s) to match 2\%  of Ethereum's total Ethhash (150TH/s)~\cite{crypto51}. So, at the moment, it is not possible to rent sufficient hashrate on NiceHash to attack these networks. However, as the markets mature, we expect to see more liquidity on them. Meanwhile, there are networks such as Expanse for which 70x of its total hashrate is regularly available for purchase on NiceHash~\cite{NiceHash}. Indeed, the less hashrate a chain has, and the more common the ability to produce that hashrate, the more likely it is that a hashrate marketplace will form with sufficient volume that a cheap double-spend is possible.  

\section{Conclusion}
\label{Conclusion}
In this work, we have shown that the low-cost of a single double-spend attack can be amplified by the threat of retaliation, inducing a no-attack equilibrium. We first modified the economic analysis developed by Budish \cite{Budish}, obtaining a similar conclusion as that drawn in this paper, namely that double-spending  attacks can be relatively cheap, and in our analysis, even free under certain conditions.
We then proposed a defense to such attacks, showing how a small cost of attack can be amplified greatly if the victim has the same capabilities as the attacker, and can counterattack the double spend. 

We  have also reported on a number of double-spend attacks that have been empirically observed. This empirical evidence suggests that attacks on  some chains are in fact quite cheap, as some were done for almost no gain.
We have only begun to see any evidence of the counterattack defense, but as the markets for hashrate power continue to mature, we expect sophisticated actors to increase their readiness to defend themselves in the case of a double-spend attack.

We close the paper with a discussion of some future directions for research. First, the disincentives for mining operators to launch double-spend attacks, coming from ownership of specialized equipment and cryptocurrency, could come under threat in the presence of a liquid  derivatives market. A mining operator can have a legitimate reason to  hedge a highly exposed  position through taking a short position. But the larger the short position,  the greater the operator's incentive to  attack the network despite being invested. Liquid derivatives markets are an explicit goal of Ethereum's ``decentralized finance'' (DeFi) movement, and there are significant Bitcoin lenders such as Genesis \cite{genesis} and Bitcoin derivatives markets such as BitMex \cite{bitmex}.

Second, a single attacker (or cartel of attackers) could double-spend several victims at once. This seems to provide a new advantage for the attacker, because the attacker would have  more at stake than each individual victim, and thus  more to lose by quitting in the War of Attrition (whereas before, the single victim had more to lose). For this reason, it may be useful to study the possibility that multiple victims could coordinate to defend against a single attacker, in addition to the possibility that bystander miners may behave strategically to protect their mining rewards.

We have shown that considering a new kind of behavior in the context of double-spend attacks can give a novel outcome. We anticipate that broader viewpoints in modeling the behavior of agents in the context of other blockchain attacks can similarly lead to new insights.

\section*{Acknowledgments} The authors would like to thank James P. Lovejoy, Eric Budish, Jacob Leshno, Mark Nesbitt, Jonathan Zittrain, James Mickens, Yiling Chen, Tadge Dryja, Nic Carter, and David Vorick for helpful discussions. This work is supported by two generous gifts to the Center for Research on Computation and Society at Harvard University, funders of the MIT Digital Currency Initiative, and NSF grant NSF CCF-15-09178. The first author was supported in part by the Ethereum Foundation.

\printbibliography

@online{KYC,
  author = {Adeyanju, Craig},
  title = {What Crypto Exchanges Do to Comply With KYC, AML and CFT Regulations},
  year = 2019,
  month = 05,
  url = {https://cointelegraph.com/news/what-crypto-exchanges-do-to-comply-with-kyc-aml-and-cft-regulations},
  lastaccessed = {2020-02-10}
}

@online{Consensys-Centralization,
  author = {Sui, Danning and Ricci, Saulo and Pfeffer, Johannes},
  title = {Are Miners Centralized? A Look into Mining Pools},
  year = 2018,
  month = 05,
  url = {https://media.consensys.net/are-miners-centralized-a-look-into-mining-pools-b594425411dc},
  lastaccessed = {2020-02-10}
}

@online{BIP50,
  author = {Andresen, Gavin},
  title = {BIP 50: March 2013 Chain Fork Post-Mortem},
  year = 2013,
  month = 03,
  url = {https://github.com/bitcoin/bips/blob/master/bip-0050.mediawiki},
  lastaccessed = {2020-02-10}
}

@online{Oligopoly,
  author = {Nick Arnosti and S. Matthew Weinberg},
  title = {Bitcoin: A Natural Oligopoly},
  year = 2018,
  month = 11,
  url = {https://arxiv.org/abs/1811.08572},
  lastaccessed = {2020-02-10}
}

@online{BCH-Difficulty,
  title = {Bit Info Charts: Bitcoin Hashrate Vs BCH Hashrate},
  url = {https://bitinfocharts.com/comparison/hashrate-btc-bch.html},
  lastaccessed = {2020-02-10}
}

@online{Bitcoin-Difficulty,
  key={Bitcoin Difficulty},
  title = {Bitcoin Hashrate Vs Difficulty},
  url = {https://btc.com/stats/diff},
  lastaccessed = {2020-02-10}
}

@online{CZ-Reorg,
  author = {Changpeng Zhao},
  title = {How should I handle blockchain forks in my DApp?},
  year = 2019,
  month = 05,
  url = {https://twitter.com/cz_binance/status/1125996194734399488},
  lastaccessed = {2020-02-10}
}

@online{Bitcoin-Pool-Centralization,
  title = {Pool Distribution},
  publisher = {BTC.com},
  url = {https://btc.com/stats/pool},
  lastaccessed = {2020-02-10}
}

@online{Largest-Bitcoin-Txns,
  title = {Largest Bitcoin Transactions by USD Value},
  publisher = {Blockchair},
  url = {https://blockchair.com/bitcoin/transactions?s=output_total_usd(desc)#},
  lastaccessed = {2020-02-10}
}

@inproceedings{Bonneau,
  title= {Why buy when you can rent? Bribery attacks on Bitcoin consensus} ,
  url={http://www.jbonneau.com/doc/B16a-BITCOIN-why_buy_when_you_can_rent.pdf},
  booktitle="BITCOIN '16: Proceedings of the 3\textsuperscript{rd} Workshop on Bitcoin and Blockchain Research",
  author="Joseph Bonneau",
  month=02,
  location="Barbados",
  year="2016",
}

@online{Budish,
  author = {Budish, Eric},
  title = {The Economic Limits of Bitcoin and the Blockchain},
  year = 2018,
  month = 06,
  url = {https://faculty.chicagobooth.edu/eric.budish/research/Economic-Limits-Bitcoin-Blockchain.pdf},
  lastaccessed = {2020-02-10}
}

@inproceedings{Instability,
  author = {Carlsten, Miles and Kalodner, Harry and Weinberg, S. Matthew and Narayanan, Arvind},
  title = {On the Instability of Bitcoin Without the Block Reward},
  year = {2016},
  url = {https://doi.org/10.1145/2976749.2978408},
  booktitle = {Proceedings of the 2016 ACM SIGSAC Conference on Computer and Communications Security},
  pages = {154–167},
  series = {CCS ’16}
}

@online{Carter-Settlement,
  author = {Carter, Nic},
  title = {It’s the settlement assurances, stupid: How to evaluate blockchains},
  year = 2019,
  month = 07,
  url = {https://medium.com/@nic__carter/its-the-settlement-assurances-stupid-5dcd1c3f4e41},
  lastaccessed = {2020-02-10}
}

@online{Day,
  author = {Allen Day and Evgeny Medvedev},
  title = {Ethereum in BigQuery: a Public Dataset for smart contract analytics},
  year = 2018,
  month = 08,
  url = {https://cloud.google.com/blog/products/data-analytics/ethereum-bigquery-public-dataset-smart-contract-analytics},
  lastaccessed = {2020-02-10}
}

@article{VOD,
  author = {Andreas Diekmann},
  title ={Volunteer's Dilemma},
  journal = {Journal of Conflict Resolution},
  volume = {29},
  number = {4},
  pages = {605-610},
  year = {1985},
  URL = {https://doi.org/10.1177/0022002785029004003}
}

@online{Top-ETH-Miners,
  title = {Ethereum Top 25 Miners by Blocks},
  url = {https://etherscan.io/stat/miner?range=14&blocktype=blocks},
  lastaccessed = {2020-02-10}
}

@article{Selfish,
  author = {Eyal, Ittay and Sirer, Emin G\"{u}n},
  title = {Majority is Not Enough: Bitcoin Mining is Vulnerable},
  year = {2018},
  issue_date = {June 2018},
  volume = {61},
  number = {7},
  url = {https://doi.org/10.1145/3212998},
  journal = {Commun. ACM},
  month = jun,
  pages = {95–102},
}

@inproceedings{Energy-Equilibria,
  author = {Fiat, Amos and Karlin, Anna and Koutsoupias, Elias and Papadimitriou, Christos},
  title = {Energy Equilibria in Proof-of-Work Mining},
  year = {2019},
  url = {https://doi.org/10.1145/3328526.3329630},
  booktitle = {Proceedings of the 2019 ACM Conference on Economics and Computation},
  pages = {489–502},
  series = {EC ’19}
}

@book{Fudenberg-Tirole-Book,
  title = {Game Theory},
  author = {Drew Fudenberg and Jean Tirole},
  year = {1991}, 
  publisher = {MIT Press},
}

@inproceedings{Mind-Mining,
  author = {Goren, Guy and Spiegelman, Alexander},
  title = {Mind the Mining},
  year = {2019},
  url = {https://doi.org/10.1145/3328526.3329566},
  booktitle = {Proceedings of the 2019 ACM Conference on Economics and Computation},
  pages = {475–487},
  series = {EC ’19}
}

@online{Hasu,
  author = {Hasu and James Prestwich and Brandon Curtis},
  title = {A Model for Bitcoin’s Security and the Declining Block Subsidy},
  year = 2019,
  month = 10,
  url = {https://uncommoncore.co/wp-content/uploads/2019/10/A-model-for-Bitcoins-security-and-the-declining-block-subsidy-v1.02.pdf},
  lastaccessed = {2020-02-10}
}

@online{Pay-To-Win,
  author = {Aljosha Judmayer and Nicholas Stifter and Alexei Zamyatin and Itay Tsabary and Ittay Eyal and Peter Gazi and Sarah Meiklejohn and Edgar Weippl},
  title = {Pay-To-Win: Incentive Attacks on Proof-of-Work Cryptocurrencies},
  year = 2019,
  month = 07,
  url = {https://eprint.iacr.org/2019/775.pdf},
  lastaccessed = {2020-02-10}
}

@article{Oceanic,
  author = {Nikos Leonardos and Stefanos Leonardos and Georgios Piliouras},
  title = {Oceanic Games: Centralization Risks and Incentives in Blockchain Mining},
  year = 2019,
  month = 05,
  journal = {International Conference on Mathematical Research for Blockchain Economy (MARBLE)},
  url = {https://arxiv.org/pdf/1904.02368.pdf},
}

@InProceedings{Whale,
  author = {Kevin Liao and Jonathan Katz},
  title = {Incentivizing Blockchain Forks via Whale Transactions},
  year = 2017,
  booktitle="Financial Cryptography and Data Security",
  pages="264--279",
  doi = {10.1007/978-3-319-70278-0_17},
  url = {https://www.cs.umd.edu/~jkatz/papers/whale-txs.pdf},
}

@online{BTG-Attacked,
  author = {James Lovejoy},
  title = {Bitcoin Gold (BTG) was 51\% attacked},
  year = 2020,
  month = 01,
  url = {https://gist.github.com/metalicjames/71321570a105940529e709651d0a9765},
  lastaccessed = {2020-02-10}
}

@online{VTC-Attacked,
  author = {James Lovejoy},
  title = {Vertcoin (VTC) was 51\% attacked},
  year = 2019,
  month = 12,
  url = {https://gist.github.com/metalicjames/f2acdb9ef448ec5298173b36c7c54133},
  lastaccessed = {2020-02-10}
}

@online{EXP-Attacked,
  author = {James Lovejoy},
  title = {Expanse (EXP) was 51\% attacked},
  year = 2019,
  month = 07,
  url = {https://gist.github.com/metalicjames/01222049f95f85df8c0eb253de54848b},
  lastaccessed = {2020-02-10}
}

@online{LCC-Attacked,
  author = {James Lovejoy},
  title = {Litecoin Cash (LCC) was 51\% attacked},
  year = 2019,
  month = 07,
  url = {https://gist.github.com/metalicjames/82a49f8afa87334f929881e55ad4ffd7},
  lastaccessed = {2020-02-10}
}

@online{double-spend-bitcointalk,
  author = {macbook-air},
  title = {A successful DOUBLE SPEND US\$10000 against OKPAY this morning},
  year = 2013,
  month = 03,
  url = {https://bitcointalk.org/index.php?topic=152348.0},
  lastaccessed = {2020-02-10}
}

@online{Nakamoto,
  author = {Nakamoto, Satoshi},
  title = {Bitcoin: A Peer-to-Peer Electronic Cash System},
  year = 2008,
  month = 10,
  url = {https://bitcoin.org/bitcoin.pdf},
  lastaccessed = {2020-02-10}
}

@online{ETC-Attacked,
  author = {Mark Nesbitt},
  title = {Deep Chain Reorganization Detected on Ethereum Classic (ETC)},
  year = 2019,
  month = 01,
  url = {https://blog.coinbase.com/ethereum-classic-etc-is-currently-being-51-attacked-33be13ce32de},
  lastaccessed = {2020-02-10}
}

@online{BTG-Verge-Attacked,
  author = {Charlie Osborne},
  title = {Bitcoin Gold suffers double-spend attacks, \$17.5 million lost},
  year = 2018,
  month = 05,
  url = {https://www.zdnet.com/article/bitcoin-gold-hit-with-double-spend-attacks-18-million-lost/},
  lastaccessed = {2020-02-10}
}

@online{Rosenfeld-DS,
  author = {Meni Rosenfeld},
  title = {Analysis of Hashrate-Based Double Spending},
  year = 2014,
  month = 02,
  url = {https://arxiv.org/abs/1402.2009},
  lastaccessed = {2020-02-10}
}

@InProceedings{Opt-Selfish,
  author={Sapirshtein, Ayelet and Sompolinsky, Yonatan and Zohar, Aviv},
  title={Optimal Selfish Mining Strategies in Bitcoin},
  booktitle={Financial Cryptography and Data Security},
  year = 2016 ,
  pages = {515--532},
  url = {https://fc16.ifca.ai/preproceedings/30_Sapirshtein.pdf}
}

@InProceedings{Maynard-Smith-74,
  author={J. Maynard Smith},
  title={The Theory of Games and the Evolution of Animal Conflicts},
  booktitle={The Journal of Theoretical Biology},
  year = 1974,
  volume = 47,
  pages = {209--221},
}

@online{Vorick,
  author = {David Vorick},
  title = {Fundamentals of Proof of Work},
  year = 2019,
  month = 01,
  url = {https://blog.sia.tech/fundamentals-of-proof-of-work-beaa68093d2b},
  lastaccessed = {2020-02-10}
}

@InProceedings{Weesie,
  author={Jeroen Weesie},
  title={Asymmetry and Timing in the Volunteer's Dilemma},
  booktitle={The Journal of Conflict Resolution},
  year = 1993,
  month = 09,
  volume = 37,
  number = 3,
  pages = {569--590},
}

@online{NiceHash,
  title = {What is NiceHash and how it works?},
  url = {https://www.nicehash.com/support/general-help/nicehash-service/what-is-nicehash-and-how-it-works},
  lastaccessed = {2020-02-10}
}

@online{ghash,
  title={Bitcoin Mining Pool Ghash.io DDos-ed in Response to threat of 51\% attack?},
  year=2014,
  month=06,
  url = {https://www.ccn.com/bitcoin-mining-pool-ghash-io-ddos-ed-response-51-attack/},
  lastaccessed = {2020-02-10}
}

@online{Zencash-Attacked,
  title = {ZenCash Statement On double-spend attack},
  year = 2018,
  month = 06,
  url = {https://blog.horizen.global/zencash-statement-on-double-spend-attack/},
  lastaccessed = {2020-02-10}
}

@online{crypto51,
  title = {Crypto51: PoW 51\% Attack Cost},
  url = {https://www.crypto51.app/},
  lastaccessed = {2020-02-10}
}

@online{Auer,
  author = {Ralph Auer},
  title = {Beyond the doomsday economics of “proof-ofwork” in cryptocurrencies},
  year = 2019,
  month = 01,
  url = {https://www.bis.org/publ/work765.pdf},
  lastaccessed = {2020-02-10}
}

@online{fees,
  title = {Bitcoin Transaction Fees},
  url = {https://btc.com/stats/fee},
  lastaccessed = {2020-02-10}
}

@online{wsj-autonomous,
  author = {Steven Russolillo and Eun-Young Jeong},
  title = {Exchanges Are Getting Hacked Because It’s Easy},
  year = 2018,
  month = 07,
  url = {https://www.wsj.com/articles/why-cryptocurrency-exchange-hacks-keep-happening-1531656000},
  lastaccessed = {2020-02-10}
}

@online{genesis,
  author = {Joeri Cant},
  title = {Genesis Capital Crypto Lending Firm Reports 870m In New Originations In Q3},
  year = 2019,
  month = 10,
  url = {https://cointelegraph.com/news/genesis-capital-crypto-lending-firm-reports-870m-in-new-originations-in-q3},
  lastaccessed = {2020-02-10}
}

@online{bitmex,
  author = {Horus Hughes},
  title = {Bitcoin Price Targets 10k As BitMex Open Interest Soars To 1.5B},
  year = 2020,
  month = 02,
  url = {https://cointelegraph.com/news/bitcoin-price-targets-10k-as-bitmex-open-interest-soars-to-15b},
  lastaccessed = {2020-02-10}
}

@online{honeylemon,
  title = {Honey Lemon Cloudmining Marketplace},
  url = {https://honeylemon.market/},
  lastaccessed = {2020-02-10}
}

@online{btg_iskra,
  author = {Edward Iskra},
  title = {Double Spend Attacks on Exchanges},
  year = 2018,
  month = 05,
  url = {https://forum.bitcoingold.org/t/double-spend-attacks-on-exchanges/1362},
  lastaccessed = {2020-02-10}
}

@online{btg_iskra_responding,
  author = {Edward Iskra},
  title = {Responding to Attacks},
  year = 2018,
  month = 05,
  url = {https://bitcoingold.org/responding-to-attacks/},
  lastaccessed = {2020-02-10}
}

\appendix
\section{Mining Pool Concentration}
Table \ref{table:btc-eth-pools} shows the percent of total hashrate owned by the top mining pools on Bitcoin and Ethereum as of February 2020. Note that just a few such pools collectively own over 50\% of the hashrate in both Bitcoin and in Ethereum. 

\begin{table}[ht]
\centering
	\begin{tabular}{ll | ll}
		\toprule
		BTC &  & ETH &   \\
		\midrule
        Pool Name & Share & Pool Name & Share  \\
		\midrule
        BTC.com	& 16.4\% & Spark Pool & 32.8\%\\
        F2Pool	& 13.9\%& Ethermine & 21.0\%\\
        Poolin	& 12.9\%& F2Pool & 11.1\%\\
        AntPool	& 11.8\%& NanoPool & 7.9\%\\
        SlushPool   & 7.2\%& Zhizhu.top & 4.1\%\\
		\bottomrule
		Total & 62.2\% & Total & 76.9\%
	\end{tabular}
	\caption{Bitcoin and Ethereum mining pools, ranked by percent of total BTC (ETH) blocks found Feb 2019 - Feb 2020 
 \protect\cite{Bitcoin-Pool-Centralization,Top-ETH-Miners}.}
\label{table:btc-eth-pools}
\end{table}

\section{Large Bitcoin and Ethereum Transactions}
\begin{figure}
\centering
\begin{minipage}{.5\textwidth}
  \centering
  \includegraphics[width=0.99\linewidth]{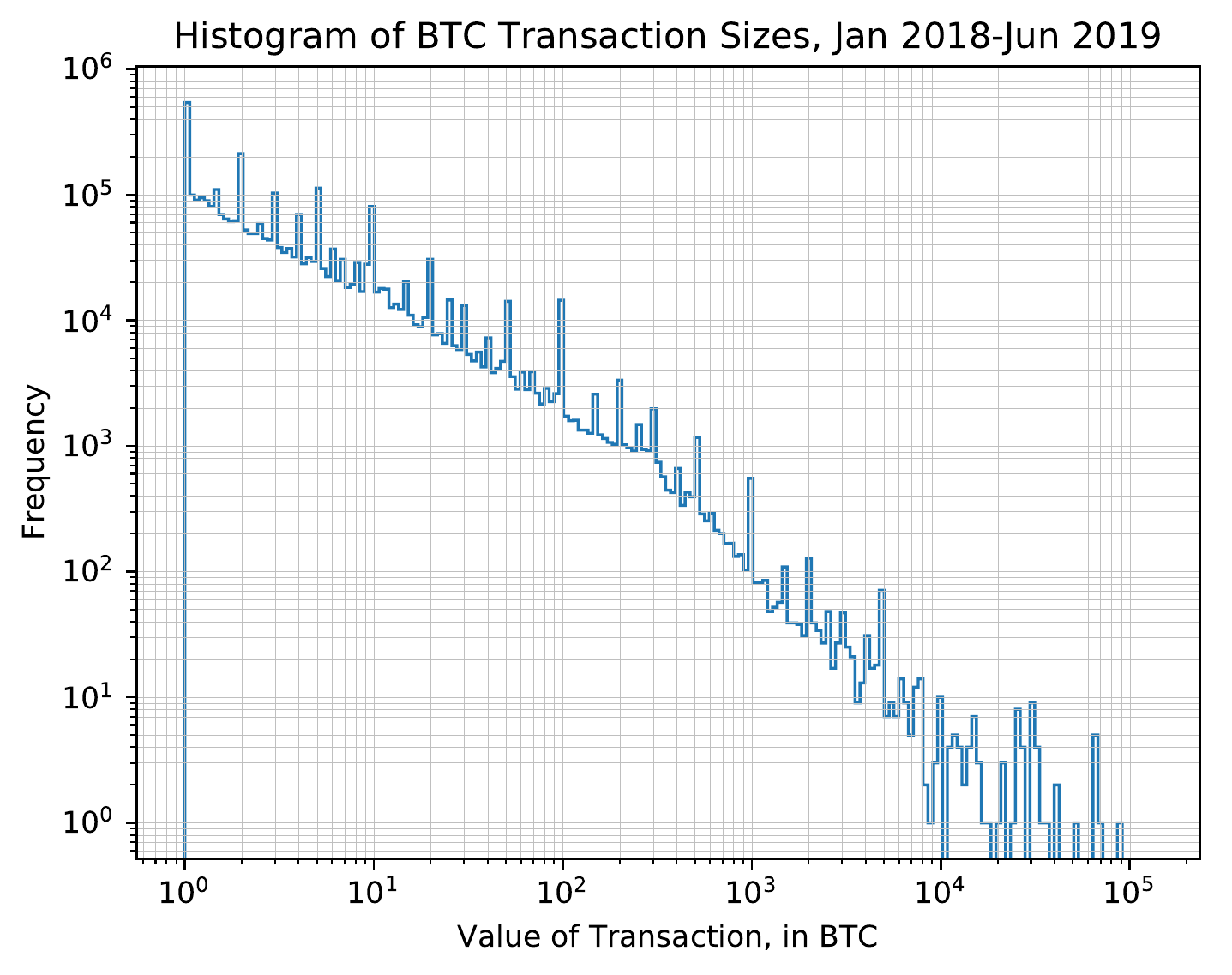}
  \caption{Histogram of BTC Transaction Sizes, Jan 2018 - Jun 2019}
  \label{fig:btc-txn-freq}
\end{minipage}%
\begin{minipage}{.5\textwidth}
  \centering
  \includegraphics[width=0.99\linewidth]{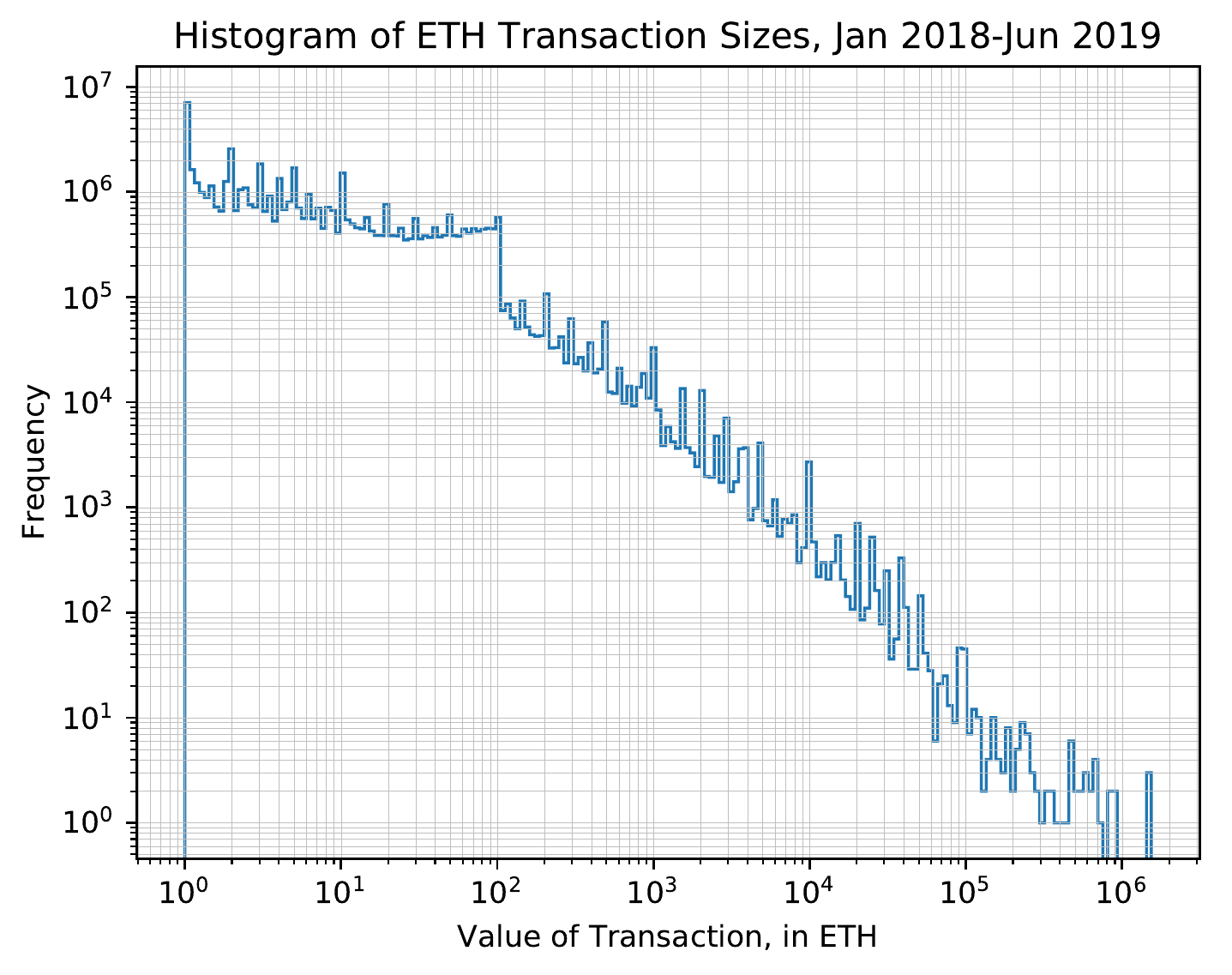}
  \caption{Histogram of ETH Transaction Sizes, Jan 2018 - Jun 2019}
  \label{fig:eth-txn-freq}
\end{minipage}
\end{figure}

Figures \ref{fig:btc-txn-freq} and \ref{fig:eth-txn-freq} show the distribution of Bitcoin and Ethereum transaction sizes, respectively, from Jan 2018 - June 2019. 
These plots were generated using data acquired by querying the Google BigQuery database that contains full-chain data on both Bitcoin and Ethereum (see Day and Medvedev (2018) \cite{Day}).
Estimating the transaction size in Bitcoin is challenging because of Bitcoin's UTXO model. For each transaction, the quantity plotted is the value of the smallest output, making this a conservative estimate of  the value of a transaction. 
See Table \ref{table:big-btc-txns} for the largest Bitcoin transactions recorded as of September 2019. 

\begin{table}[ht]
\centering
	\begin{tabular}{lcccc}
		\toprule
        Block &	Date & BTC (K) & USD (B) & Fee \\
		\midrule
        578328	& 05.29.19	& 157.46	& 1.37	& 107\\
        578327	& 05.29.19	& 144.08	& 1.25	& 19\\
        576358	& 05.16.19	& 122.80	& 1.00	& 11\\
        593467  & 09.06.19  & 94.50     & 1.02  & 700\\
        577462	& 05.23.19	& 121.80	& 0.93	& 10\\
        499773	& 12.17.17	& 48.50	    & 0.89	& 15\\
        503030	& 01.07.18	& 51.04	    & 0.85	& 227\\
        498970	& 12.12.17	& 48.50	    & 0.80	& 178\\
        574898	& 05.06.19	& 125.80	& 0.73	& 9\\
        575024	& 05.07.19	& 124.30	& 0.72	& 9\\
		\bottomrule
	\end{tabular}
	\caption{The largest Bitcoin transactions as of Sept 06 2019 by USD Value, the data coming from Blockchair \protect\cite{Largest-Bitcoin-Txns}. BTC column is in thousands of Bitcoins, USD column is in Billions of USD, and Fee is in USD}
\label{table:big-btc-txns}
\end{table}

\end{document}